\documentclass[12pt]{iopart}
\expandafter\let\csname equation*\endcsname\relax
\expandafter\let\csname endequation*\endcsname\relax
\usepackage{iopams}
\usepackage{amsmath}
\usepackage{graphicx}
\usepackage{amssymb}
\usepackage{color,soul}
\usepackage[T1]{fontenc}
\usepackage{ae,aecompl}

\begin{document}

\title{Computing with volatile memristors: An application of {\it non-pinched} hysteresis}

\begin{abstract}
The possibility of in-memory computing with volatile memristive devices,
namely, memristors requiring a power source to sustain their memory, is
demonstrated. We have adopted a hysteretic graphene-based field emission
structure as a prototype of volatile memristor, which is characterized by a
non-pinched hysteresis loop. Memristive model of the structure is developed
and used to simulate a polymorphic circuit implementing in-memory computing
gates such as the material implication. Specific regions of parameter space
realizing useful logic functions are identified. Our results are
applicable to other realizations of volatile memory devices.
\end{abstract}

\author{Y.~V.~Pershin}
\ead{pershin@physics.sc.edu}
\address{Department of Physics and Astronomy, University of South Carolina, Columbia, South Carolina
29208, USA}
\address{Nikolaev Institute of Inorganic Chemistry SB RAS, Novosibirsk 630090, Russia}

\author{S.~N.~Shevchenko}
\address{B.~I.~Verkin Institute for Low Temperature Physics and Engineering, Kharkov
61103, Ukraine}
\address{V.~N.~Karazin Kharkov National University, Kharkov 61022, Ukraine}

\maketitle

\section{Introduction}

\label{sec:Intro}

Currently, there is a strong interest in in-memory computing concept. In
particular, there are expectations that in-memory computing architectures
may help to overcome the Von Neumann bottleneck problem~\cite{Backus78a} of
conventional computers and thus provide us with better computing machines.
Memristive~\cite{chua76a} (memory resistive) and memcapacitive~\cite%
{diventra09a} (memory capacitive) elements that combine information
processing and storage functionalities in simple device structures of
nanoscale dimensions have received a great deal of attention in the context
of in-memory computing (memcomputing~\cite{diventra13a})
paradigm. In fact, the material implication gate was demonstrated
experimentally with non-volatile memristive devices several years ago~\cite%
{Borghetti10}. This idea has been further developed and reviewed in a number
of papers \cite{yang2013memristive,Kvatinsky14a,Kvatinsky14b,traversa14a,Linn15a,Linn15b,pershin15a}.

While there is a wide variety of physical systems with memory~\cite%
{Pershin11}, it is generally agreed that the non-volatile memory devices are
the most suitable candidates for in-memory computing, and for good reason.
In this paper, however, we explore a different route to in-memory computing
based on \textit{volatile} memristive devices. It is shown that, in
principle, simple circuits of volatile memristors can provide some useful
logic functions. Here, we do not aim to develop a practical
in-memory computing architecture, but rather present a proof of concept
application of volatile memristors. Eventually, it may find its own
application niche.

To make our description physically based, in this paper we consider the
hysteretic behavior of carbon-based field emitters~\cite{Eletskii10,
Fedoseeva15, Li15, Kleshch15, Gorodetskiy16}. For concreteness, we
have chosen a hysteretic graphene-based field emission structure~\cite%
{Kleshch15} as a prototype of volatile memristor. The memory effect in such
a structure is attributed to a field-induced detachment of a portion of graphene
sheet from substrate~\cite{Kleshch15}. As in this system the minimum voltage required to induce
an OFF to ON transition $V_{\mathrm{ON}}$ is larger than that needed for the
transition from ON to OFF, $V_{\mathrm{OFF}}$, there is a voltage interval
where the structure remembers its state (defined by the history of voltage
applied).

Thus, there are two main results reported in this paper: (i) the memristive
model of graphene field emitters, and (ii) realization of in-memory
computing gates based on such devices. Accordingly, this paper is organized
as follows. We develop a memristive model of hysteretic graphene-based field
emitters in Sec.~\ref{sec:model}. In particular, in the first part of this
Section we formulate general equations of the model, while in the second
part (that may be skipped by those
readers who are not interested in model details), we formulate the model
parameters based on our understanding of
 physical processes associated with graphene detachment
from substrate. In Sec.~\ref{sec:logic}, an implementation
of logic gates based on volatile memristors is explored. We conclude in
Sec.~\ref{sec:conclusion} with a summary of our study.

\section{Memristive model of graphene field emitters}

\label{sec:model}

In this Section we develop a memristive model of graphene field emitters~%
\cite{Kleshch15} showing that such devices can be classified as first-order
voltage-controlled memristive systems. Our model is well suited for the
description of experimental results, as it captures both the switching
dynamics and physics of field emission. We emphasize that the suggested
memristive model can be adopted for the description of other nanomechanical
systems with memory including those~\cite{Nieminen02,Sun16} that do not
 require high voltages for their operation.

\begin{figure}[b]
\centering{\includegraphics[width=8cm]{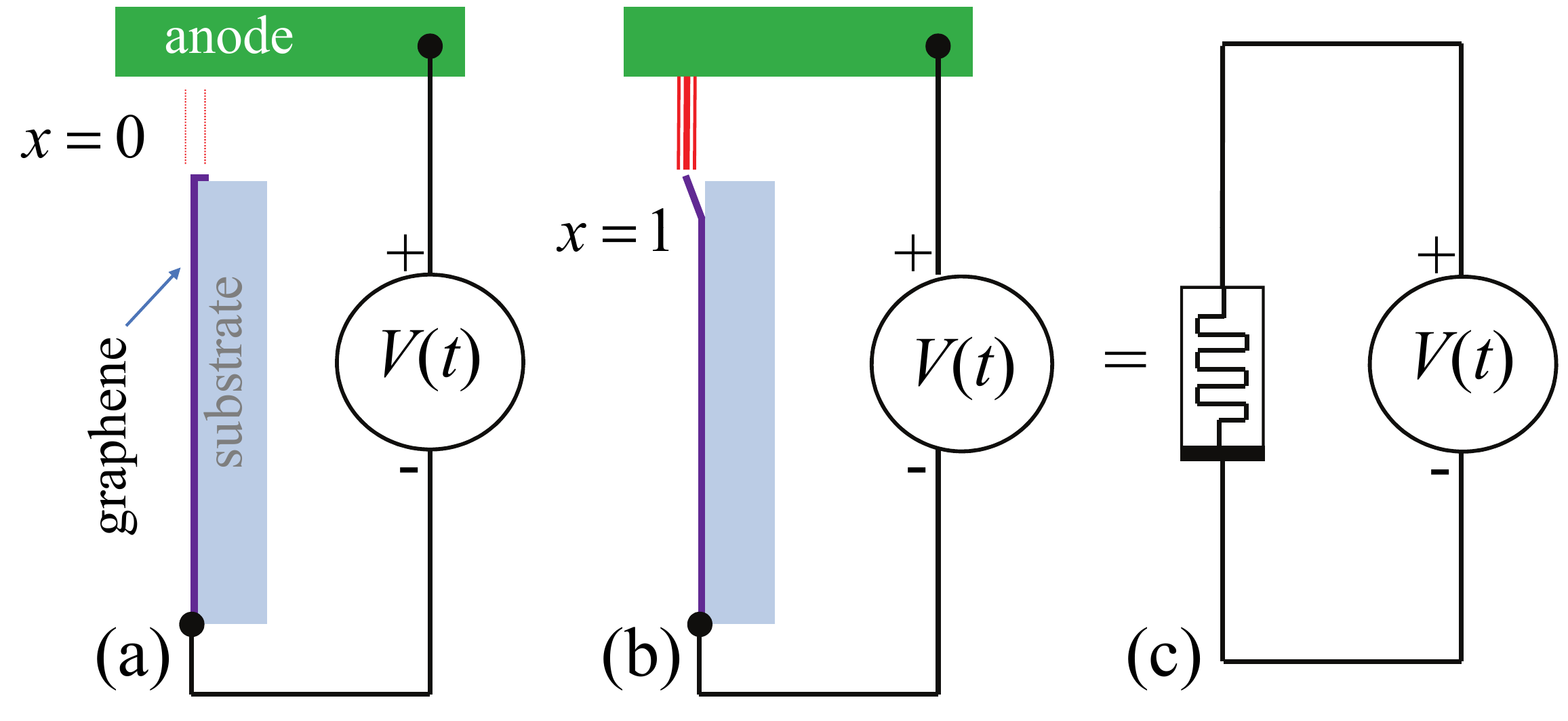}}
\caption{(a) and (b): Schematic representation of low- and high-current
states of the graphene field emitter: (a) the low-current state (the edge is
attached to substrate, $x=0$) and (b) the high-current state (the edge is
detached/standing, $x=1$). Both states are stable at $V_{\mathrm{OFF}}<V<V_{\mathrm{ON}}$. (c)
Memristive circuit model of circuits in (a) and (b).}
\label{fig1}
\end{figure}

\subsection{Memristive model}

In a recent experiment~\cite{Kleshch15} a strong hysteresis in
current-voltage characteristics of field emission from the edge of graphene
on SiO$_{2}$ was observed. This behavior was explained by a field-assisted
local detachment of the graphene edge from the substrate (for a schematic
illustration see Fig.~\ref{fig1}). In particular, it was demonstrated that
when the system is subjected to an increasing voltage $V$ from 0 to a
maximum value, there is a rapid increase in the current at a certain $%
V_{switch}$ (in what follows, denoted by $V_{\mathrm{ON}}$). On the way back, a
current drop is observed at $V_{\mathrm{OFF}}<V_{\mathrm{ON}}$ such that $V_{\mathrm{ON}}/V_{\mathrm{OFF}}%
\approx 7$. Importantly, in the hysteretic region (ranging from $V_{\mathrm{OFF}}$ to
$V_{\mathrm{ON}}$) the current is stable in the sense that the system can stay
arbitrary long in one of two (in some samples, many) possible current states. Thus, the
memory of such field emitters can be classified as long-term and volatile
(the memory is lost at small $V$ including $V=0$). A similar memory effect
in field emission from graphene was also observed in our own in-house
experiments \cite{in-house}.

In order to describe the hysteretic field emission from graphene, we use the
formalism of memristive devices developed by Chua and Kang~\cite{chua76a}.
According to the definition, an $N$-order voltage-controlled memristive
system is given by
\begin{eqnarray}
I(t) &=&R_{M}^{-1}(\mathbf{x},V,t)V(t),  \label{eq:I(t)} \\
\dot{\mathbf{x}} &=&f(\mathbf{x},V,t),  \label{eq:x_dot}
\end{eqnarray}%
where $R_{M}$ is the memristance (memory resistance), which depends on the
input voltage $V$ and vector $\mathbf{x}$ of $N$ internal state variables. The
function $f$ in Eq.~(\ref{eq:x_dot}) defines the dynamics of internal state.
Nowadays, Eqs.~(\ref{eq:I(t)})-(\ref{eq:x_dot}) are widely used to model a
broad range of emergent non-volatile memory devices~\cite{Pershin11}.
Moreover, the present authors applied Eqs.~(\ref{eq:I(t)})-(\ref{eq:x_dot})
to field emission from carbon nanotubes \cite{Gorodetskiy16}.

It is natural to select the internal state variable $x$ as $x=L_{p}/L_{tot}$%
, where $L_{p}$ is the length of detached (standing) portion of the edge,
and $L_{tot}$ is the edge length. Two limit cases (completely attached, $x=0$%
, and detached, $x=1$, edges) are schematically depicted in Fig.~\ref{fig1}.
Generally, $x$ can take any intermediate value between 0 and 1. To formulate
the memristive model of graphene field emitters, we assume that the current
in $x=0$ and $x=1$ states can be described by the Fowler-Nordheim law~\cite%
{FN28a}. Note that this assumption is in agreement with experimental
observations~\cite{Kleshch15}.

The total current can be written as a sum of currents through the attached
and detached regions of the edge:
\begin{equation}
I=\left( 1-x\right) I_{\mathrm{OFF}}+xI_{\mathrm{ON}},  \label{I}
\end{equation}%
where $I_{\mathrm{OFF}}$ and $I_{\mathrm{ON}}$ are the total emission currents at $x=0$ and $%
x=1$, respectively. $I_{\mathrm{OFF}}$ and $I_{\mathrm{ON}}$ are represented using the
Fowler-Nordheim law as
\begin{equation}
I_{\mathrm{OFF(ONN)}}=A_{\mathrm{OFF(ON)}}V^{2}e^{-\frac{B_{\mathrm{OFF(ON)}}%
}{V}}.  \label{eq:FN}
\end{equation}%
Here, $A_{\mathrm{OFF(ON)}}$ and $B_{\mathrm{OFF(ON)}}$ are constants discussed in Subsec. \ref%
{sec2b}.

\begin{figure}[t]
\centering{\includegraphics[width=8cm]{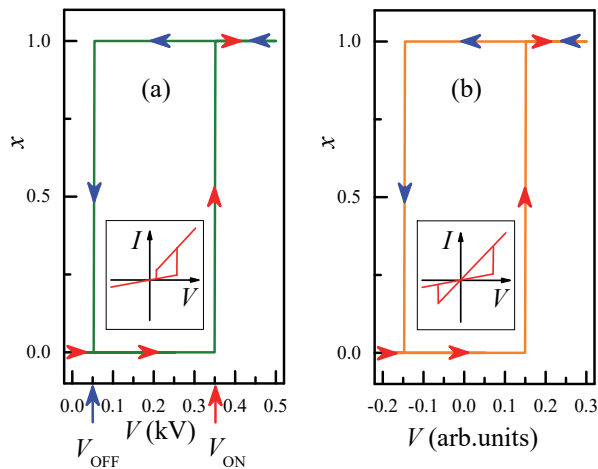}}
\caption{Hysteretic curves for the internal state variable $x$ of (a)
volatile (graphene field emitter) and (b) hypothetical nonvolatile
memristor. Insets demonstrate respective non-pinched and pinched hysteretic $%
I-V$ curves.}
\label{fig2}
\end{figure}

In order to reproduce experimental results~\cite{Kleshch15}, it is
sufficient to select the function $f$ in the Eq.~(\ref{eq:x_dot}) as
\begin{equation}
f(V)=\left\{
\begin{array}{cc}
\gamma & \;\textnormal{if}\;V\geq V_{\mathrm{ON}} \\
-\gamma & \;\textnormal{if}\;V\leq V_{\mathrm{OFF}} \\
0 & \textnormal{otherwise}%
\end{array}%
,\right.  \label{eq:f(V,M)}
\end{equation}%
where $\gamma >0$ is the rate of change of $x$. In fact, the function $f$ defined by Eq.~(%
\ref{eq:f(V,M)}) can describe both types of memristors: non-volatile and
volatile. Assuming a positive $V_{\mathrm{ON}}$, the memristor type is defined by inequalities
\begin{eqnarray}
V_{\mathrm{ON}} &>&V_{\mathrm{OFF}}\geq0\text{  :  volatile,} \\
V_{\mathrm{ON}} &>&0>V_{\mathrm{OFF}}\text{   :  non-volatile.}  \notag
\end{eqnarray}%
Figure~\ref{fig2} schematically shows examples of the dynamics of $x$ in a
volatile memristor (such as the graphene field emitter), Fig. \ref{fig2}(a), and in a hypothetical
non-volatile memristor, Fig. \ref{fig2}(b), subjected to a periodic quasistatic waveform voltage.

\begin{figure}[t]
\centering{\includegraphics[width=8cm]{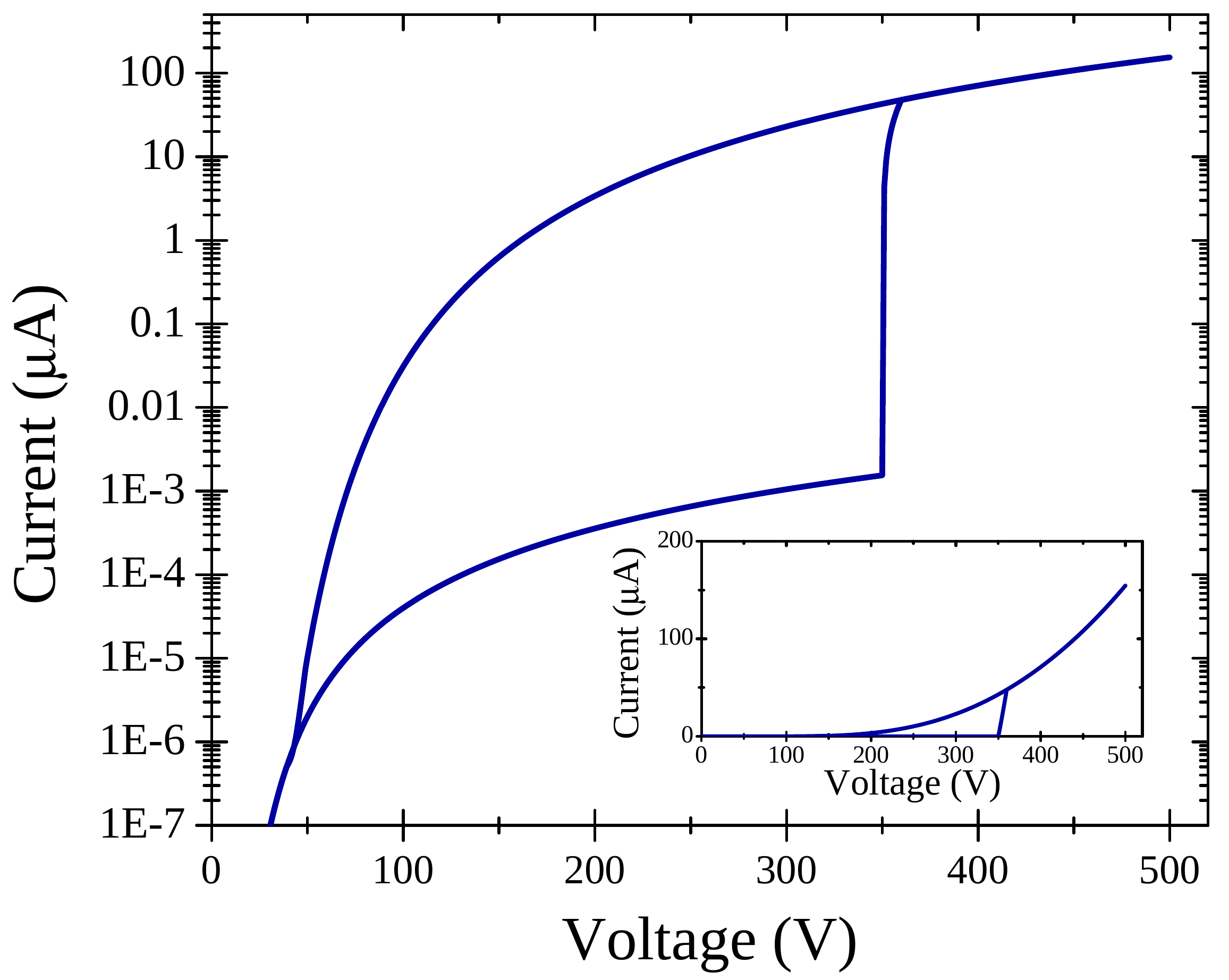}}
\caption{$I-V$ curve of the graphene field emitter found using Eqs.~(\protect
\ref{eq:I(t)})-(\protect\ref{eq:f(V,M)}) with the following set of parameter
values: $V_{\mathrm{OFF}}=50$ V, $V_{\mathrm{ON}}=350$ V, $A_{\mathrm{ON}}=2.32\cdot 10^{-9}$ A/V$^{2}$%
, $B_{\mathrm{ON}}=662.2$ V, $A_{\mathrm{OFF}}=1.99\cdot 10^{-14}$ A/V$^{2}$, $B_{\mathrm{OFF}}=160.6$
V, $\protect\gamma T=100$, where $T$ is the voltage period. Inset: the same
curve shown in the linear scale.}
\label{fig3}
\end{figure}

A calculated $I-V$ curve of graphene field emitter subjected to a triangular waveform
voltage is shown in Fig.~\ref{fig3}. We emphasize that our volatile
memristor exhibits a non-pinched hysteresis.

\subsection{Physical basis of the model}

\label{sec2b}

Here, we briefly discuss the expressions for the model parameters $A_{\mathrm{OFF(ON)%
}}$ and $B_{\mathrm{OFF(ON)}}$.

Consider the field emission from a graphene-based cathode, as presented in
Fig.~\ref{fig1}. The potential difference $V(t)$ between the cathode and
anode results in the electric field $E=\beta V/D$, where $D$ is the
distance between the electrodes and $\beta$ is the form factor. Then the
current is described by the Fowler-Nordheim formula \cite{Eletskii10,
Sheshin01, in-house}%
\begin{eqnarray}
I(V) &=&AV^{2}\exp \left( -B/V\right) ,  \label{FN} \\
A &=&\frac{e^{3}}{16\pi ^{2}\hbar }\frac{1}{\varphi }\left( \frac{\beta }{D}%
\right) ^{2}S,\text{ \ \ }B=\frac{4\sqrt{2m}}{3e\hbar }\varphi ^{3/2}\left(
\frac{\beta }{D}\right) ^{-1},  \notag
\end{eqnarray}%
where $e$ and $m$ are the electron charge and mass, $\hbar $ is the Planck constant,
 $S$ is the effective emitting surface, and $\varphi =4.8$ eV
is the work function.

In Fig.~\ref{fig1}, the situation (a) corresponds to the
graphene sheet entirely attached to the substrate, while in the case (b), the
edge of the sheet is detached. Following the arguments put forward in Refs.~\cite{Kleshch15, in-house},
we believe that the main effect is likely associated with the change in
the form factor $\beta$ and effective emitting surface $S$.
Introducing $\beta _{\mathrm{OFF(ON)}}$ and $S_{\mathrm{OFF(ON)}}$ for the low- and high-current
states, the model parameters are defined as $A_{\mathrm{OFF(ON)}}\equiv A(S_{\mathrm{OFF(ON)}},\beta _{\mathrm{OFF(ON)}})$ and $B_{\mathrm{OFF(ON)}}\equiv B(S_{\mathrm{OFF(ON)}},\beta _{\mathrm{OFF(ON)}})$.
An intermediate situation is described by the superposition state,
Eq.~(\ref{I}).

\section{Logic gates}

\label{sec:logic}

\begin{figure}[t]
\centering{\includegraphics[width=7cm]{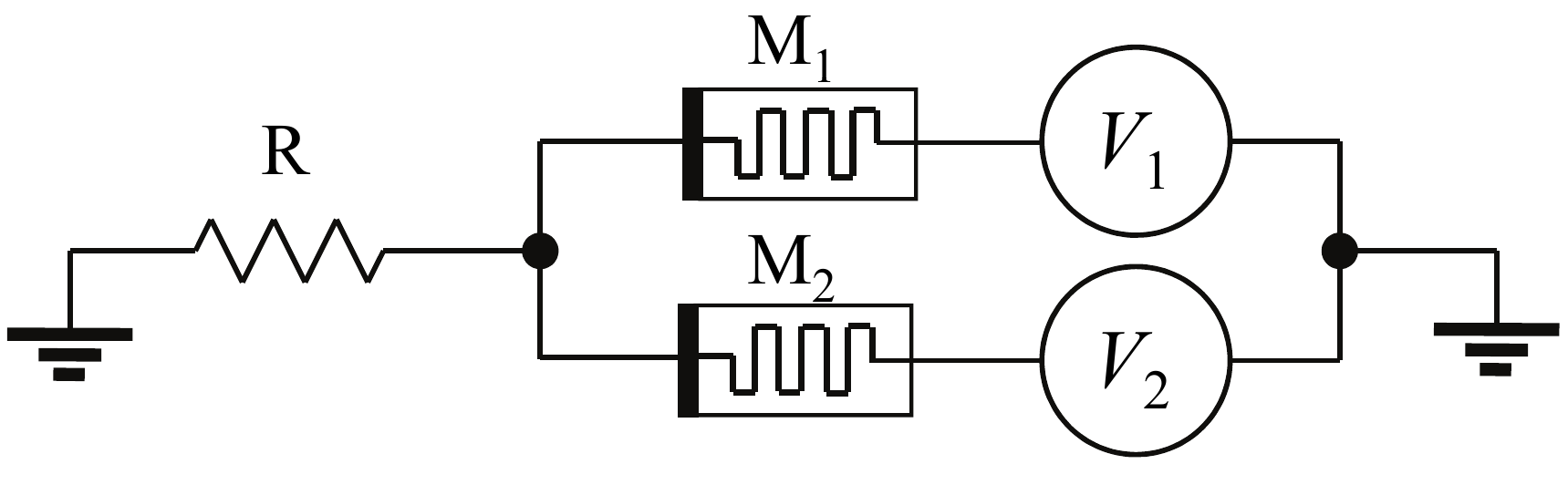}}
\caption{In-memory computing circuit considered in this work. The circuit
combines two memristors M$_i$, resistor R and two voltage sources.}
\label{fig4}
\end{figure}

\subsection{Circuit and calculation of the operation code}

The possibility of in-memory computing with volatile memristors is
investigated employing Fig.~\ref{fig4} circuit, which is similar to the
circuit used in the demonstration of the material implication with
non-volatile memristors~\cite{Borghetti10}. In what follows this circuit
is simulated based on the Kirchhoff's circuit laws equation for $V_R(t)$
\begin{equation}
\frac{ V_{1}-V_R(t)}{R_{M,1}} +\frac{ V_{2}-V_R(t)}{R_{M,2}} =\frac{%
V_{R}(t)}{R},  \label{Kirchhoff}
\end{equation}
which is supplemented by Eqs. (\ref{eq:I(t)}), (\ref{eq:x_dot}) for the dynamics of
memristances $R_{M,1}$ and $R_{M,2}$. In Eq. (\ref{Kirchhoff}), $V_R(t)$ is the
voltage across R.

\begin{table}[bh]
\caption{Codes~\protect\cite{pershin15a} of logic operations calculated
according to Eq.~(\protect\ref{code}). These codes are defined with respect
to different pairs of initial states of M$_{1}$ and M$_{2}$ and can describe
the final state of any device of interest (in our case, M$_{1}$ or M$_{2}$).
For more information, see the text and Ref.~\cite{pershin15a}.}%
\renewcommand{\arraystretch}{1.3} \centering
\begin{tabular}{|c|c||c|c||c|c|}
\hline
set to 0 & 0 & XOR & 6 & copy M$_1$ & 12 \\ \hline
NOR & 1 & NAND & 7 & IMP$_2$ & 13 \\ \hline
NOT(IMP$_2$) & 2 & AND & 8 & OR & 14 \\ \hline
NOT M$_1$ & 3 & NOT(XOR) & 9 & set to 1 & 15 \\ \hline
NOT(IMP$_1$) & 4 & copy M$_2$ & 10 &  &  \\ \hline
NOT M$_2$ & 5 & IMP$_1$ & 11 &  &  \\ \hline
\end{tabular}%
\label{codes}
\end{table}

\begin{figure*}[t]
\centering{(a)\includegraphics[width=6cm]{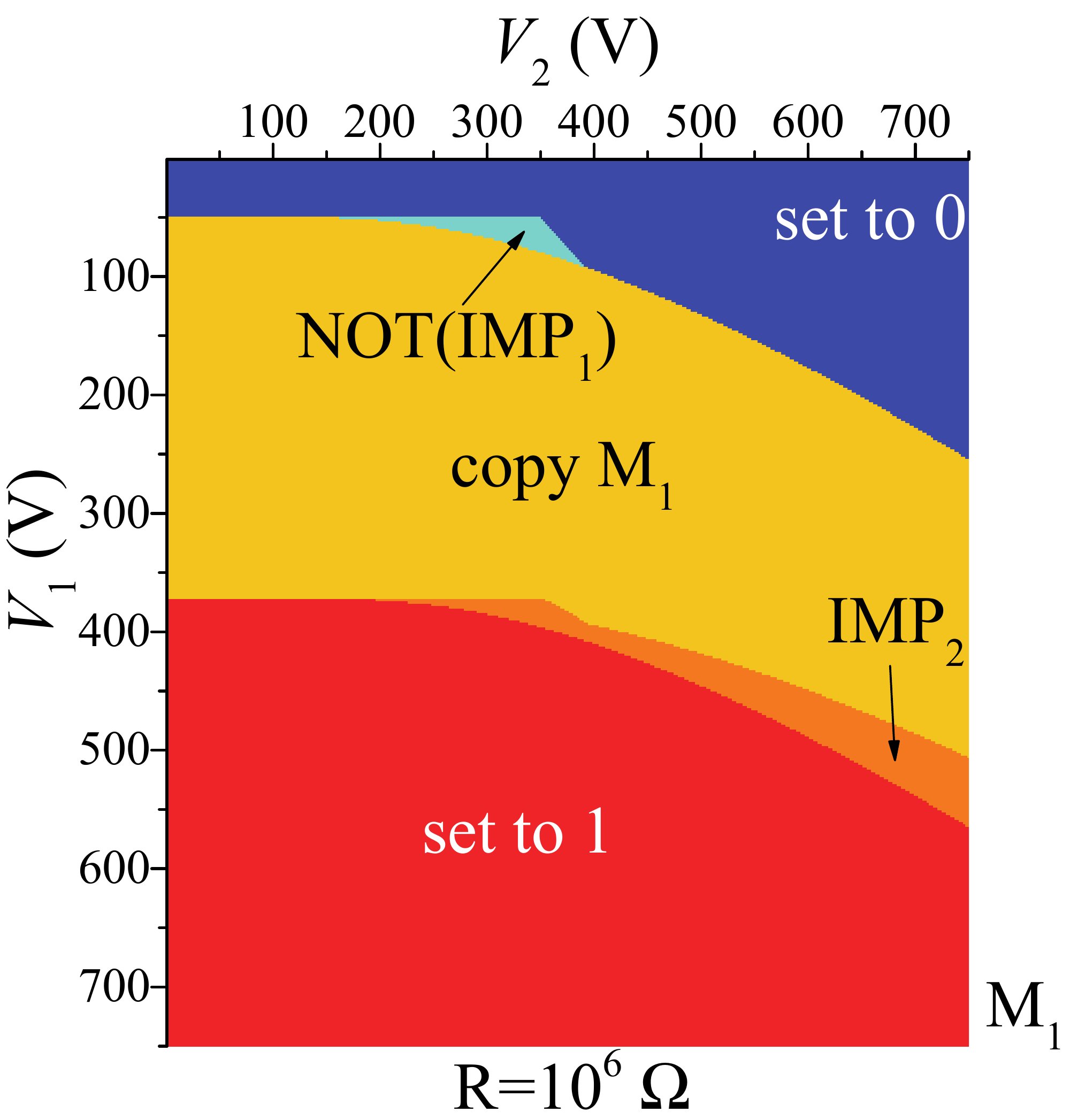} \hspace{1cm}
(b) \includegraphics[width=6cm]{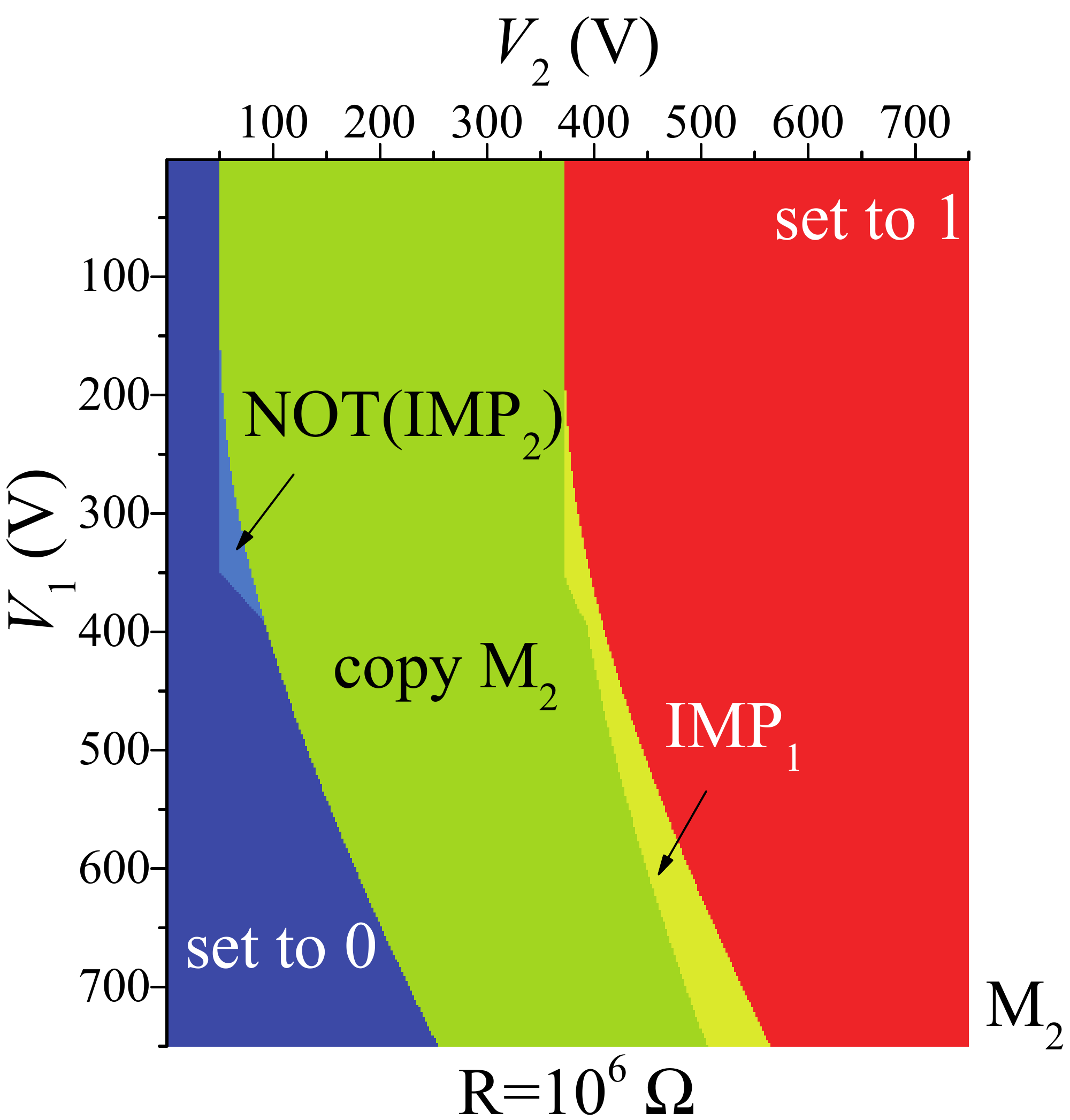}}
\caption{Operation type as a function of applied voltages calculated using
Fig.~\protect\ref{fig4} circuit with $R=10^6$ $\Omega$. The final states of M%
$_1$ and M$_2$ hold the logic function outputs presented in (a) and (b),
respectively. These plots were obtained with the same parameters of M$_1$ and M$%
_2$ as in Fig.~\protect\ref{fig3}.}
\label{fig5}
\end{figure*}

\begin{figure*}[tb]
\centering{(a)\includegraphics[width=6cm]{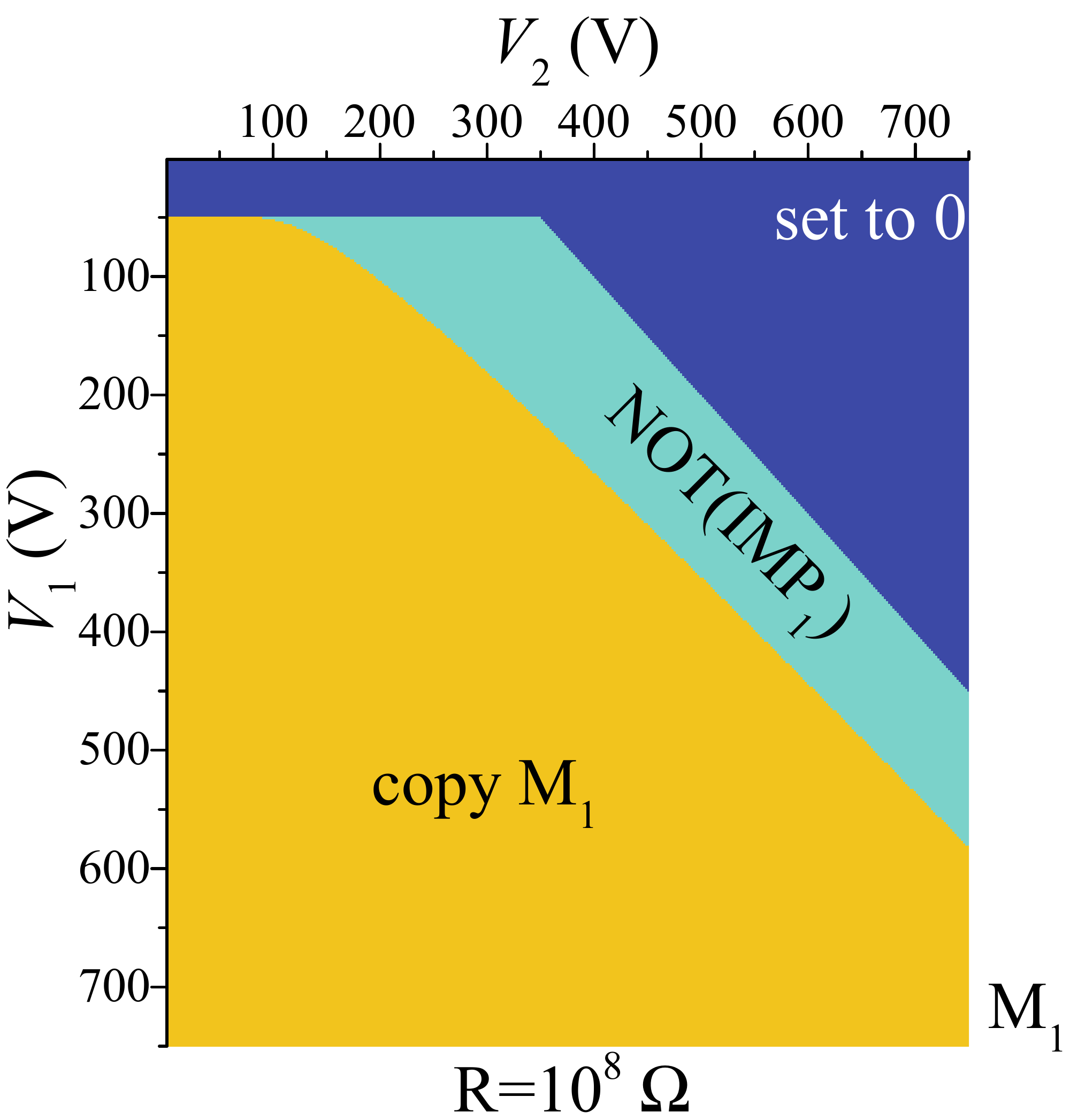} \hspace{1cm}
(b) \includegraphics[width=6cm]{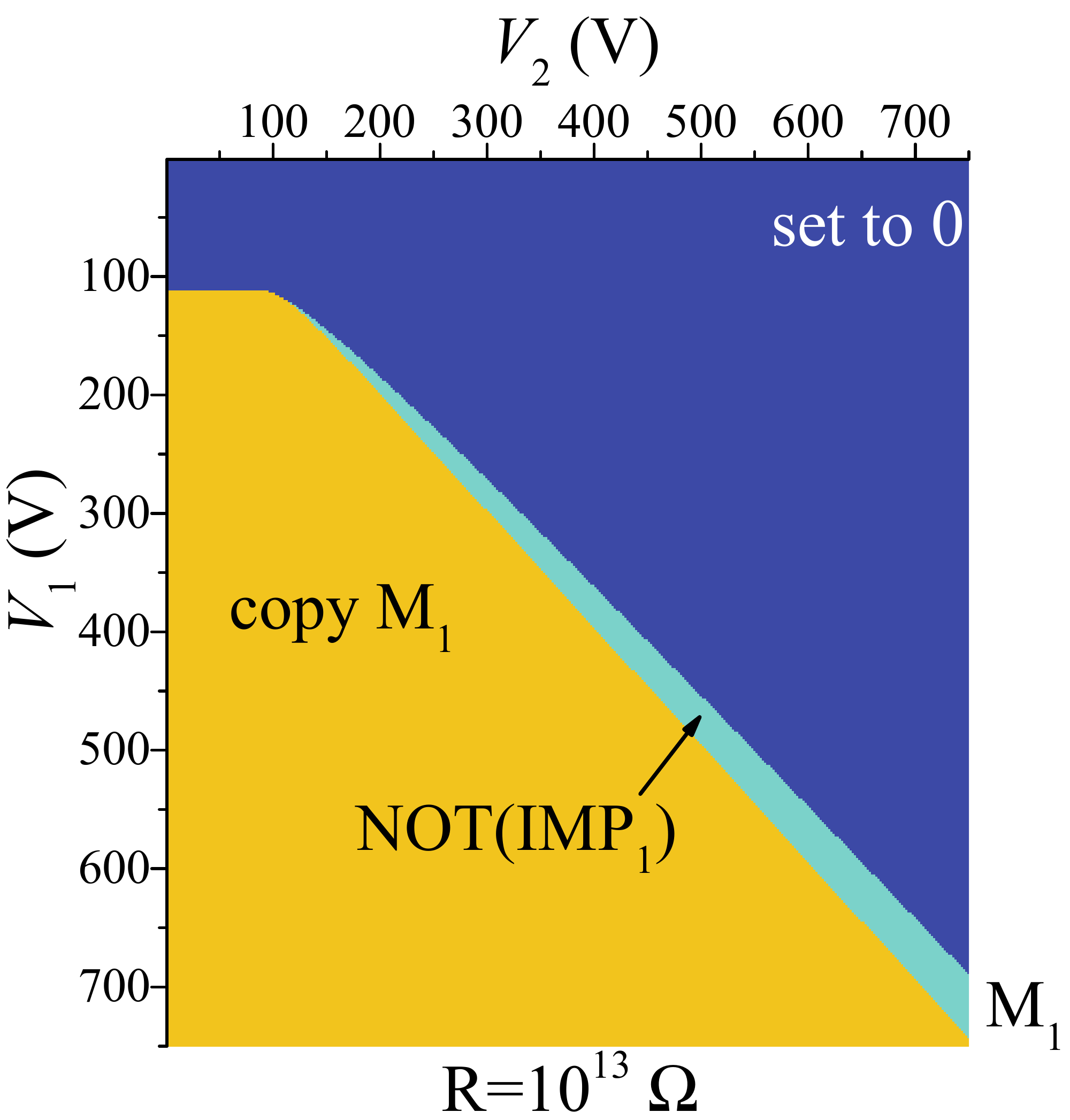}}
\caption{Operation type as a function of applied voltages calculated using
Fig.~\protect\ref{fig4} circuit with $R=10^{8}$ $\Omega $ and $10^{13}$ $%
\Omega $. These plots were obtained with the same parameters of M$_{1}$ and M%
$_{2}$ as in Fig.~\protect\ref{fig3}.}
\label{fig6}
\end{figure*}

Following Ref.~\cite{pershin15a}, we analyze the simulations results
calculating a numerical code that can be associated with a specific logic
operation. Taking $w_{i}=1,2,4,8$ as weights for the input
combinations (0,0), (0,1), (1,0) and (1,1), the numerical code is calculated
as a weighted sum of the final state of a selected memristor,
\begin{equation}
\mathnormal{code}=\sum\limits_{i=1}^{4}w_{i}b_{ij}^{f},  \label{code}
\end{equation}%
where $b_{ij}^{f}$ is the final state ($0$ or $1$) of the device of interest
$j$ (in our case, M$_{1}$ or M$_{2}$) for $i$-th input combination $(0,0)$, $%
(0,1)$, $(1,0)$ or $(1,1)$ that correspond to $i=1,2,3,4$. Table~\ref{codes}
summarizes logic functions for all possible code values. In this Table the standard
notations are used for the logic functions, e.g., NOT is the logical
negation, IMP is the material implication (in particular, IMP$_{1}$ is M$%
_{1}\rightarrow $M$_{2}$), etc. In our numerical simulations of Fig.~\ref%
{fig4} circuit, we have encountered the following operation codes: $0$, $2$,
$4$, $10$-$13$, $15$.

\subsection{Diagrams of logic operations}

Figs.~\ref{fig5} and \ref{fig6} show some selected results of our
simulations. In order to obtain each point of these plots, we simulated the
dynamics of Fig.~\ref{fig4} circuit for all possible pairs of initial states
of M$_{1}$ and M$_{2}$ subjected to $V_{1}$ and $V_{2}$. The operation code
was found with Eq.~(\ref{code}) and interpreted based on the Table~\ref%
{codes}.

\begin{figure*}[tb]
\centering{(a)\includegraphics[width=6cm]{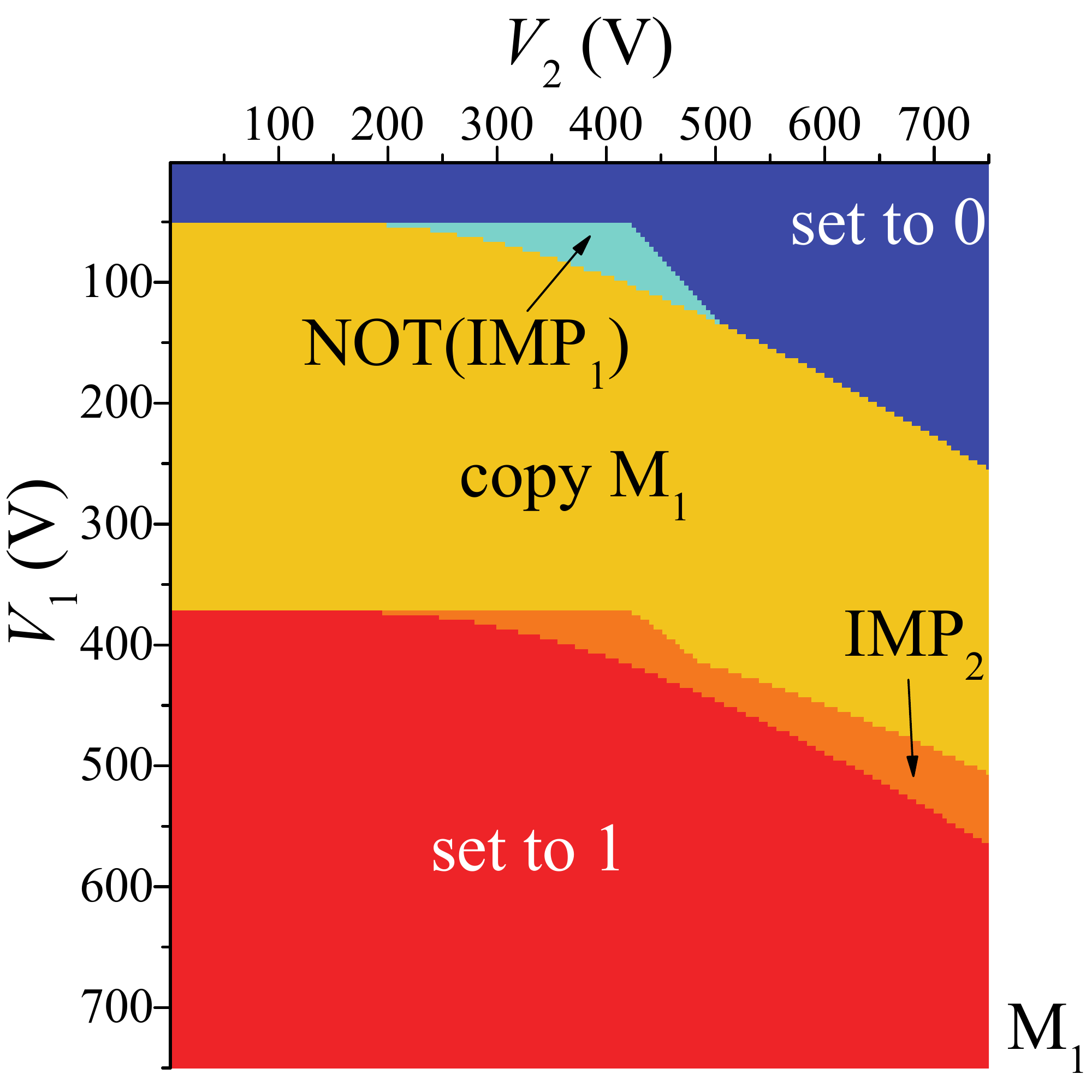} \hspace{1cm}
(b) \includegraphics[width=6cm]{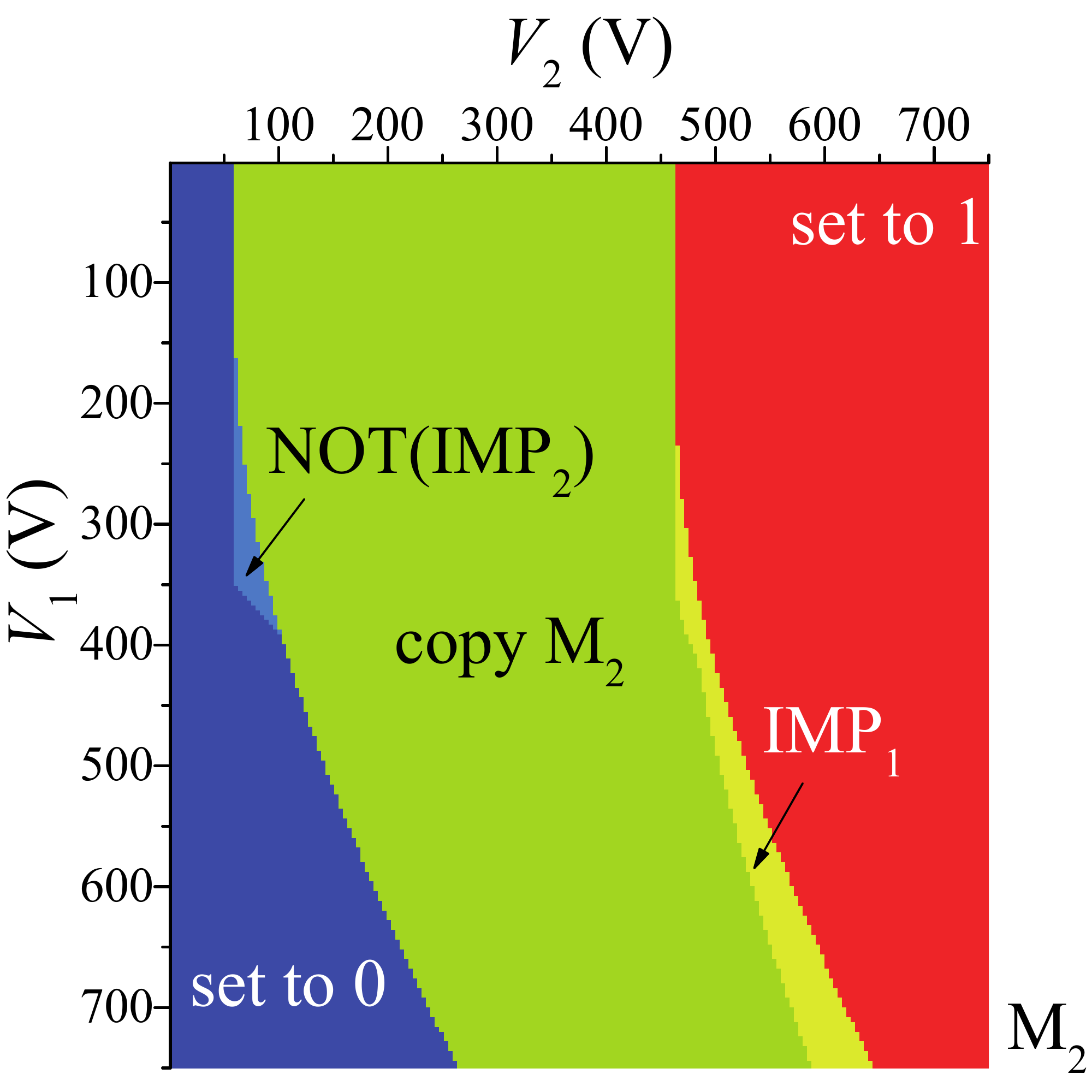}} \centering{(c)%
\includegraphics[width=6cm]{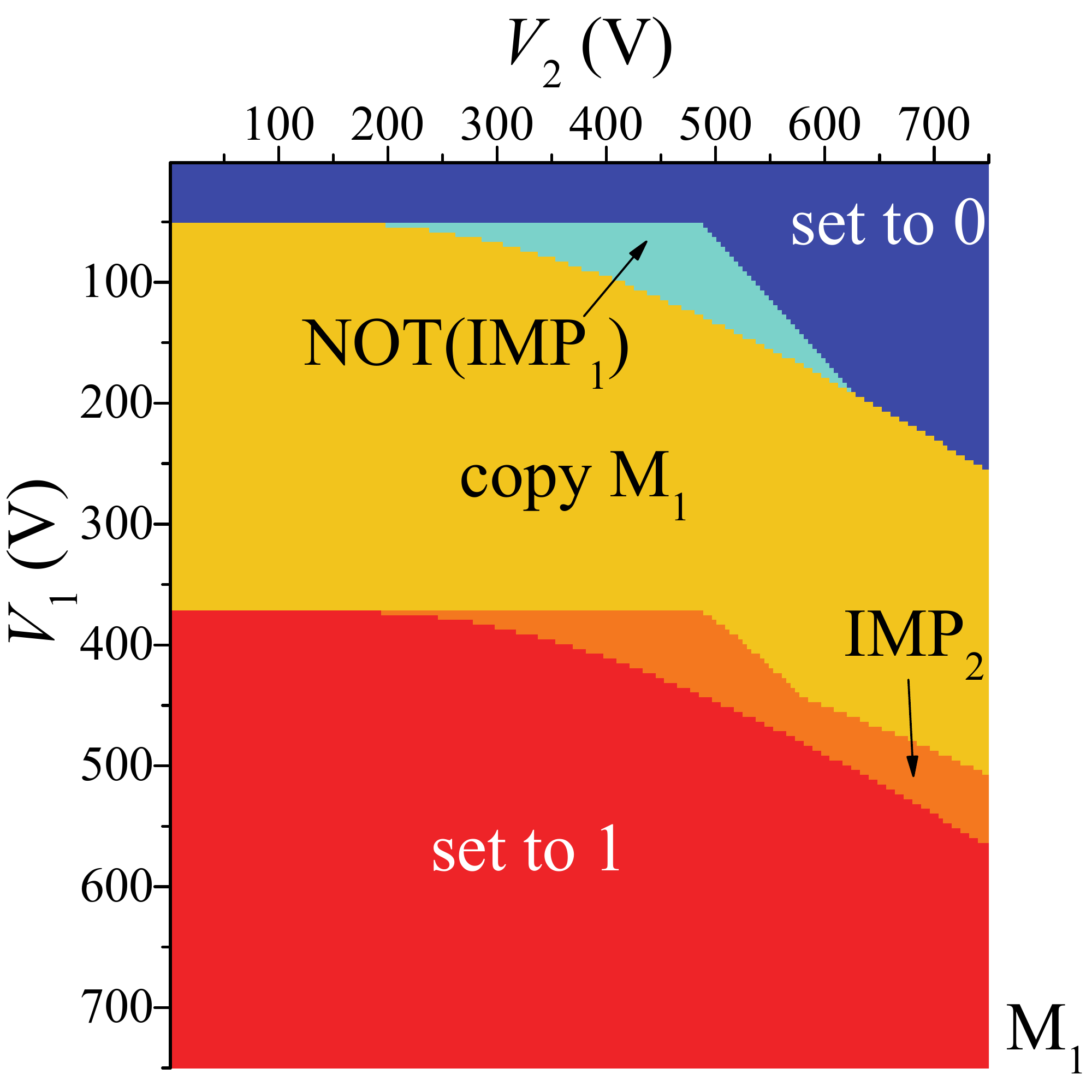} \hspace{1cm} (d) %
\includegraphics[width=6cm]{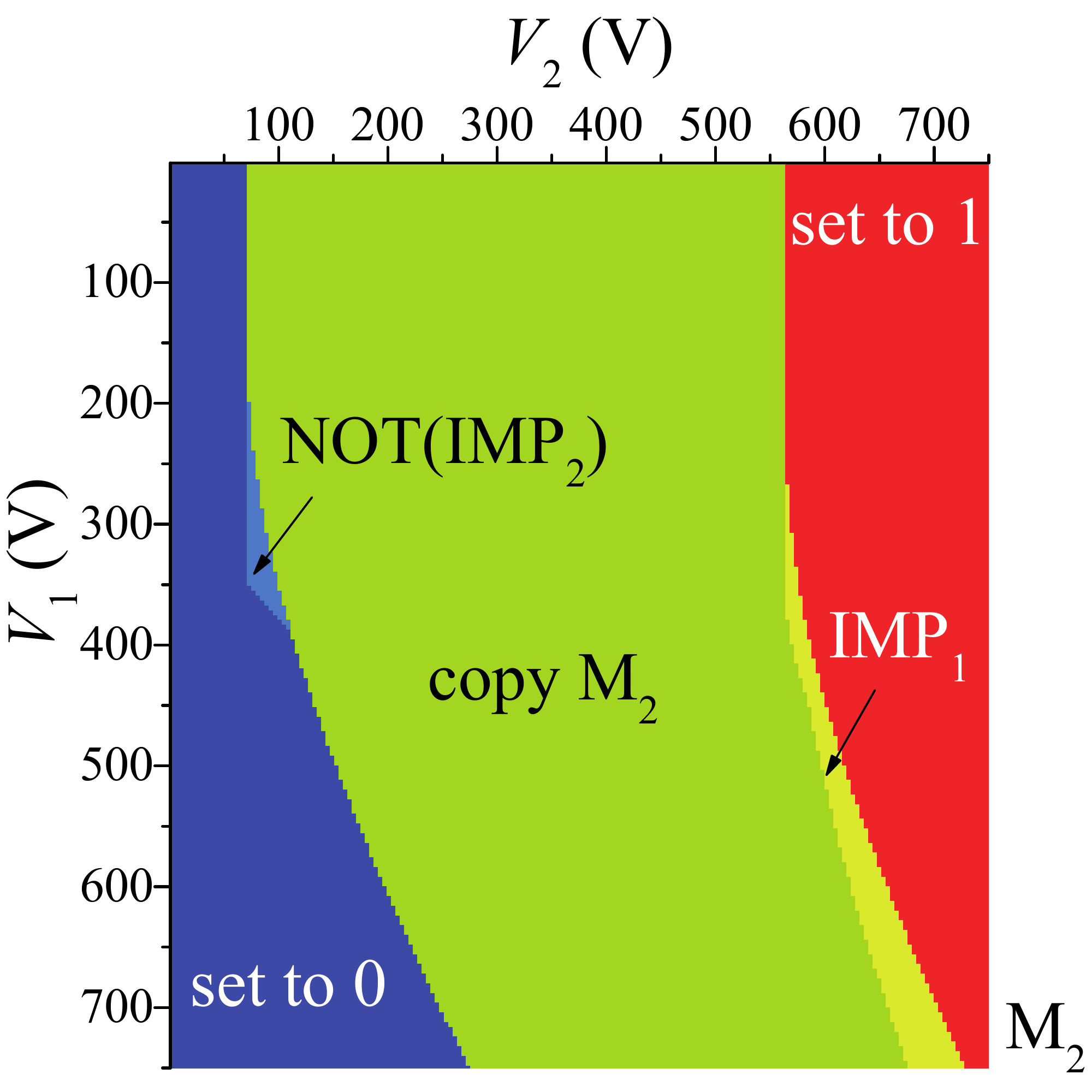}} .
\caption{Effect of variability of memristor parameters. To obtain these
plots we used $R=10^6$ $\Omega$, and higher $V_{\mathrm{ON(OFF)}}$ for M$_{2}$: $V_{\mathrm{OFF}}=60$ V
and $V_{\mathrm{ON}}=420$ V in (a) and (b), and $V_{\mathrm{OFF}}=70$ V and $V_{\mathrm{ON}}=490$ V in
(c) and (d).  All other model parameters were as in Fig.~\protect\ref{fig3}. Compare with Fig.~\protect\ref{fig5}.}
\label{figApp1}
\end{figure*}

According to Fig.~\ref{fig5}, the logic operations are symmetric for M$_{1}$
and M$_{2}$ with respect to $V_{1}=V_{2}$ line. As expected, at low voltages
applied to M$_{i}$, $x_{i}$ changes to 0, at high voltages -- to 1, and
there is also a stability region (copy to M$_{i}$). At $R=0$, the common
stability region is a square defined by the lines $V_{i}=V_{\mathrm{ON(OFF)}}$. This
square is deformed at $R>0$ (this can be seen by placing Fig.~\ref{fig5}(b)
over Fig.~\ref{fig5}(a) or vice versa). The most important voltages regions,
however, are those providing the material implication (IMP) and negation of
implication (NOT(IMP)) gates. The importance of the material implication
stems from the fact that it is a fundamental logic gate~\cite%
{whitehead1912principia}, which, together with 'set to 0' (FALSE) operation
form a computationally complete logic basis.

Fig.~\ref{fig6} shows the effect of the resistance of R on logic operations
regions. One can notice that, generally, an increase in $R$ scales the
operation regions in Fig.~\ref{fig5} (a) to higher voltages. In particular,
one can notice the disappearance of 'set to 1' regions (these regions are now beyond
the scales presented) and, in fact, an increase of the region of NOT(IMP).
This observation, actually, is of value as the proper choice of R simplifies
the experimental realization of logic gates and improves reliability.

In order to demonstrate the proposed logic gates experimentally,
one can implement, for example, the following operation
protocol. First of all, the memristors can be independently initialized by
grounding the common point of their connection with R and applying suitable voltage sequences
$V_1(t)$ and $V_2(t)$.
Next, the grounding of the connection point is released while
$V_{1}$ and $V_{2}$ are kept in the stability region of memristors
(operation codes $10$ and $12$). Third, $V_{1}$
and $V_{2}$ can be simultaneously placed into the desired operation point
and switched back into the stability region. The calculation results will be
stored in the final states of memristors.

\subsection{Parameter variability effects}

In this subsection we investigate the effect of variability of memristor
parameters on the logic functions realized with Fig.~\ref{fig4} circuit.
Specifically, we consider the operation of Fig.~\ref{fig4} circuit employing
memristors with different threshold voltages.
For this purpose, the simulations are performed using higher values of threshold voltages
of M$_{2}$ keeping all other simulation parameters as in Fig.~\ref{fig5}
simulations. Fig.~\ref{figApp1} presents two examples of such calculations
showing the diagrams found at about $20\%$ and $40\%$ higher threshold
voltages of M$_{2}$ compared to M$_{1}$.

In Fig.~\ref{figApp1}, one can notice that the diagrams for M$_{1}$ and M$%
_{2} $ are no more symmetric. At the same time, the general topologies of
diagrams are the same as these in Fig.~\ref{fig5}. Importantly, the areas of
useful logic functions for M$_{1}$ (the implication and negation of
implication) increase with an increase in $V_{\textnormal{OFF}}$ and $V_{\textnormal{ON}}$ of M$_{2}$. This
observation can be used, e.g., to achieve more stable operation of such
memristive logic gates.

\section{Conclusion}

\label{sec:conclusion}

We considered the possibility of in-memory computing (in the form of boolean
logic) based on volatile memristive devices. As a prototype of such
structures, a hysteretic graphene field emitter was adopted. A memristive
model of field emission from the graphene cathode was developed. This model
is practical for the description of real experiments.

Moreover, it was shown that simple circuits of volatile memristors can serve
as a polymorphic logic gate. Specifically, we have demonstrated that in
addition to the trivial operation set (FALSE, TRUE and hold the state) the
same circuit can implement the material implication and the negation
of implication. We expect that volatile memristors could find their own
applications, e.g., in low-level information processing circuits.

\section{Acknowledgment}

This work has been supported by the Russian Scientific Foundation grant No.
15-13-20021. The authors gratefully acknowledge fruitful discussions with A.~V.~Okotrub and D.~V.~Gorodetskiy.

\section*{References}

\bibliography{IEEEabrv,memristor}

\begin{thebibliography}{10}

\bibitem{Backus78a}
J.~Backus.
\newblock Can programming be liberated from the von {N}eumann style? {A}
  functional style and its algebra of programs.
\newblock {\em Comm. Assoc. Comp. Machin.}, 21:613--641, 1978.

\bibitem{chua76a}
L.~O. Chua and S.~M. Kang.
\newblock Memristive devices and systems.
\newblock {\em Proc. {IEEE}}, 64:209--223, 1976.

\bibitem{diventra09a}
M.~{Di Ventra}, Y.~V. Pershin, and L.~O. Chua.
\newblock Circuit elements with memory: Memristors, memcapacitors, and
  meminductors.
\newblock {\em Proc. {IEEE}}, 97(10):1717--1724, 2009.

\bibitem{diventra13a}
M.~Di~Ventra and Y.~V. Pershin.
\newblock The parallel approach.
\newblock {\em Nature Physics}, 9:200, 2013.

\bibitem{Borghetti10}
J.~Borghetti, G.~S. Snider, P.~J. Kuekes, J.~J. Yang, D.~R. Stewart, and R.~S.
  Williams.
\newblock Memristive switches enable stateful logic operations via material
  implication.
\newblock {\em Nature}, 464:873--876, 2010.

\bibitem{yang2013memristive}
J.~J. Yang, D.~B. Strukov, and D.~R. Stewart.
\newblock Memristive devices for computing.
\newblock {\em Nature nanotechnology}, 8(1):13--24, 2013.

\bibitem{Kvatinsky14a}
S.~Kvatinsky, G.~Satat, N.~Wald, E.~G. Friedman, A.~Kolodny, and U.~C. Weiser.
\newblock Memristor-based material implication (imply) logic: Design principles
  and methodologies.
\newblock {\em {IEEE} Trans. {VLSI} Syst.}, 22(10):2054--2066, 2014.

\bibitem{Kvatinsky14b}
S.~Kvatinsky, D.~Belousov, S.~Liman, G.~Satat, N.~Wald, E.~G. Friedman,
  A.~Kolodny, and U.~C. Weiser.
\newblock Magic memristor-aided logic.
\newblock {\em {IEEE} Trans. Circuits Syst. {II}}, 61(11):895--899, 2014.

\bibitem{traversa14a}
F.~L. Traversa, F.~Bonani, Y.~V. Pershin, and M.~Di Ventra.
\newblock Dynamic computing random access memory.
\newblock {\em Nanotechnology}, 25:285201, 2014.

\bibitem{Linn15a}
A.~Siemon, S.~Menzel, R.~Waser, and E.~Linn.
\newblock A complementary resistive switch-based crossbar array adder.
\newblock {\em {IEEE} J. Emerg. Sel. Topics Circuits Syst.}, 5(1):64--74, 2015.

\bibitem{Linn15b}
A.~Siemon, S.~Menzel, A.~Chattopadhyay, R.~Waser, and E.~Linn.
\newblock In-memory adder functionality in 1s1r arrays.
\newblock In {\em 2015 IEEE Int. Symp. Circ. Syst. (ISCAS)}, pages 1338--1341,
  2015.

\bibitem{pershin15a}
Y.~V. Pershin, F.~L. Traversa, and M.~Di Ventra.
\newblock Memcomputing with membrane memcapacitive systems.
\newblock {\em Nanotechnology}, 26:225201, 2015.

\bibitem{Pershin11}
Y.~V. Pershin and M.~Di~Ventra.
\newblock Memory effects in complex materials and nanoscale systems.
\newblock {\em Adv. Phys.}, 60:145, 2011.

\bibitem{Eletskii10}
A.~V. Eletskii.
\newblock Carbon nanotube-based electron field emitters.
\newblock {\em Physics-Uspekhi}, 53:863, 2010.

\bibitem{Fedoseeva15}
Y.~V. Fedoseeva, L.~G. Bulusheva, A.~V. Okotrub, M.~A. Kanygin, D.~V.
  Gorodetskiy, I.~P. Asanov, D.~V. Vyalikh, A.~P. Puzyr, and V.~S. Bondar.
\newblock Field emission luminescence of nanodiamonds deposited on the aligned
  carbon nanotube array.
\newblock {\em Sci. Rep.}, 5:9379, 2015.

\bibitem{Li15}
Y.~Li, Y.~Sun, and J.~T.~W. Yeow.
\newblock Nanotube field electron emission: principles, development, and
  applications.
\newblock {\em Nanotechnology}, 26:242001, 2015.

\bibitem{Kleshch15}
Victor~I. Kleshch, Denis~A. Bandurin, Anton~S. Orekhov, Stephen~T. Purcell, and
  Alexander~N. Obraztsov.
\newblock Edge field emission of large-area single layer graphene.
\newblock {\em Appl. Surf. Sci.}, 357, Part B:1967 -- 1974, 2015.

\bibitem{Gorodetskiy16}
D.~V. Gorodetskiy, A.~V. Gusel'nikov, S.~N. Shevchenko, M.~A. Kanygin, A.~V.
  Okotrub, and Y.~V. Pershin.
\newblock Memristive model of hysteretic field emission from carbon nanotube
  arrays.
\newblock {\em J. Nanophoton.}, 10:012524, 2016.

\bibitem{Nieminen02}
H~Nieminen, V~Ermolov, K~Nybergh, S~Silanto, and T~Ryhanen.
\newblock Microelectromechanical capacitors for rf applications.
\newblock {\em J. Micromech. Microeng.}, 12:177, 2002.

\bibitem{Sun16}
J.~Sun, M.~E. Schmidt, M.~Muruganathan, H.~M.~H. Chong, and H.~Mizuta.
\newblock Large-scale nanoelectromechanical switches based on directly
  deposited nanocrystalline graphene on insulating substrates.
\newblock {\em Nanoscale}, 8:6659, 2016.

\bibitem{in-house}
D.~V. Gorodetskiy, S.~N. Shevchenko, A.~V. Okotrub, and Y.~V. Pershin.
\newblock {\em to be published}.

\bibitem{FN28a}
R.~H. {Fowler} and L.~{Nordheim}.
\newblock {Electron Emission in Intense Electric Fields}.
\newblock {\em Proc. R. Soc. Lond. A}, 119:173--181, 1928.

\bibitem{Sheshin01}
E.~P. Sheshin.
\newblock {\em Surface structure and electron field emission properties of
  carbon materials}.
\newblock MFTI, Moscow, 2001.

\bibitem{whitehead1912principia}
A.~N. Whitehead and B.~Russell.
\newblock {\em Principia mathematica}, volume~2.
\newblock University Press, 1912.

\end{thebibliography}
\bibliographystyle{unsrt}

\end{document}